# Structure-driven analog optical control in ion-pumped $SrFeO_{3-\delta}$ thin-film devices

Alicia Ruiz-Caridad[1], Paul Nizet[1], Francesco Chiabrera[1], Xavier Vea[1], Philipp Langner[1], Alex Morata[1], Albert Tarancon[1].

*Department of Advanced Materials for Energy Catalonia Institute for Energy Research (IREC) Jardin de les Dones de Negre 1, Sant Adrià de Besòs (Barcelona) 08930, Spain*

**Abstract**

Electrochromic devices (ECDs) offer a compelling route toward low-power, non-emissive optical modulators with nonvolatile states. However, their widespread implementation is hindered by limitations in operating voltage, switching speed, color tunability, and long-term stability. Mixed ionic–electronic conductors (MIECs) provide a promising alternative platform, enabling optical modulation through ion-driven redox and structural transformations. Oxygen-based MIECs offer enhanced durability, environmental robustness, and compatibility with oxide electronics and silicon photonics, yet remain largely underexplored for electrochromic and photonic applications.

Here, we demonstrate structure-driven analog optical control in an ion-pumped $SrFeO_{3-\delta}$ thin-film device by undergoing reversible oxygen-driven phase transitions between brownmillerite and perovskite structures. Phase transition is accompanied by pronounced changes in its electronic structure and optical constants. By harnessing these ion-induced structural transformations and integrating an optically passive $Al_2O_3$ interference layer, we achieve continuous and reversible modulation of optical transmittance and color. These results provide a general framework for ion-driven analog photonic and electrochromic devices and highlight the potential of oxygen-based MIECs for next-generation ionochromic systems compatible with silicon-based photonic platforms.

## 1. Introduction

Electrochromic devices (ECDs) represent a promising class of next-generation optical modulators, in which color and optical transmittance are controlled through electrochemically driven redox processes [1]. Unlike emissive display technologies such as LEDs or OLEDs, electrochromic systems operate by modulating light absorption rather than generating light, making them more closely related to passive modulation technologies such as LCDs, which control light through electric-field-induced changes. This fundamental difference offers several advantages, including reduced energy consumption, nonvolatile optical states, and minimized blue-light emission, which is associated with long-term ocular damage [1-3]. As a result, ECDs have attracted significant interest for applications ranging from smart windows and antiglare mirrors to electronic skins or wearable devices [4-6].

Despite these advantages, achieving high-performance electrochromic devices remains challenging. Functional ECDs require low operating voltages, high optical contrast, fast and reversible switching, gradual color tuning, and long-term stability. Over the past years, extensive efforts have been devoted to exploring novel organic, inorganic, and hybrid materials for electrochromic applications [7]. Inorganic metal oxides typically offer low power consumption and high chemical stability but often suffer from slow switching kinetics and limited color tunability [8]. Metal alloys, while electrically conductive, generally exhibit low optical contrast and stability. Organic electrochromic materials and conjugated polymers can achieve fast switching and multicolor modulation; however, their scalability, long-term stability, and reliance on toxic organic solvents pose significant limitations [9]. Consequently, despite decades of research, the widespread commercialization of competitive electrochromic displays remains limited [10].

As a potential solution, mixed ionic–electronic conductors (MIECs) have emerged as attractive candidates for electrochromic devices. MIECs comprise inorganic oxides, polymers, and hybrid systems [11] . Among the these types of active materials, ceramics simultaneously conduct electronic carriers and mobile ions, such as $H^+$, $Li^+$, or $O^{2-}$, under an applied electric field [12]. Ion migration and associated redox processes can induce substantial changes in the electronic structure and optical properties of the host material. While MIECs have been extensively investigated in electrochemical energy technologies, including batteries, sensors, and solid oxide fuel cells, their potential for electrochromic and optoelectronic applications remains comparatively underexplored. Lithium- and proton-based systems have enabled reversible electrochromic modulation, tunable optical transmittance, interference color, refractive index control, and plasmonic resonances, including voltage-gated optical devices, full-color nanophotonic structures, and waveguide-integrated plasmochromic modulators [19-23]. Conventional electrochromic oxides such as $WO_3$, $V_2O_5$, NiO, and related compounds have also been extensively investigated for ion-driven modulation of optical absorption and color in both conventional electrochromic devices and integrated photonic platforms [24,25]. Despite these advances, most reported systems rely on carrier-density modulation, intercalation-driven electrochromism, plasmonic resonance shifts, or switching between discrete optical states.

In contrast, oxygen-ion-based MIECs offer several unique advantages for electrochromic operation. Compared with hydrogen- and lithium-based systems, oxygen-based oxides exhibit reduced volatility, lower sensitivity to ambient conditions, and improved compatibility with oxide electronics and fully solid-state architectures [12,13]. In addition, oxygen-ion systems are compatible with robust oxide frameworks and high-temperature processing, making them particularly attractive for integrated photonic and electronic platforms. Recent studies have provided important insights into oxygen vacancy formation, migration dynamics, and nanoscale oxygen redistribution in thin-film oxides, significantly advancing the understanding of oxygen-driven electrochemical phenomena [14,15]. In oxygen-ion-conducting and mixed ionic–electronic conducting oxides, praseodymium-doped ceria has emerged as a model system, demonstrating temporal and spatial tuning of optical constants through electrochemical oxygen exchange, as well as active modulation of visible–UV optical properties linked to oxygen non-stoichiometry and defect chemistry [16,17]. Related MIEC-based devices have further demonstrated self-holding optical actuation, highlighting the potential of solid-state oxygen redistribution for nonvolatile optical modulation [18]. More recently, oxide-based photonic systems exploiting electrically controlled oxygen stoichiometry have attracted increasing interest because of their compatibility with solid-state architectures, nonvolatile operation, and potential integration with silicon photonics [26-28]. Nevertheless, continuous analog optical modulation driven by reversible oxygen-controlled structural phase evolution in epitaxial mixed ionic–electronic conductors remains comparatively unexplored. Establishing a direct relationship between oxygen stoichiometry, structural evolution, optical constants, and perceptual color within a single ion-controlled oxide platform remains an outstanding challenge.

In this context, $SrFeO_{3-\delta}$ exhibits a reversible structural transformation between the oxygen-deficient brownmillerite phase ($SrFeO_{2.5}$) and the fully oxidized perovskite phase ($SrFeO_3$), mediated by changes in oxygen stoichiometry. The BM-PV transformation does not occur as a simple two-state transition but involves intermediate phases with progressively varying oxygen content and structural order [30]. This continuous evolution of intermediate states, rather than abrupt phase switching, is consistent with recent quantitative studies of oxygen stoichiometry control in $SrFeO_{3-\delta}$ [31]. Quantitative electrochemical measurements in $SrFeO_{3-\delta}$ demonstrate continuous tuning of oxygen stoichiometry across intermediate equilibrium states, with a marked change in behavior near $3 - \delta \approx 2.7$, consistent with the emergence of phase coexistence. Moreover, all the phases can be controlled through the application of an external voltage, which drives oxygen ion migration within the lattice. The oxygen stoichiometry was quantitatively estimated from electrochemical charge integration measurements, following the relation between transferred charge and oxygen non-stoichiometry described in the Methods section. This approach enables continuous tracking of oxygen incorporation and extraction across intermediate equilibrium states and provides a quantitative basis for correlating structural evolution with optical modulation.

As a result, the material can be reversibly tuned across a continuum of states. This phase evolution is accompanied by pronounced modifications in lattice symmetry, Fe coordination, and electronic structure, which directly impact the optical response of the material, enabling voltage-controlled modulation of its optical properties [32].

In this work, we address these challenges by demonstrating structure-driven analog optical control in an ion-pumped $SrFeO_{3-\delta}$ thin-film device. $SrFeO_{3-\delta}$ is an oxide that undergoes reversible, oxygen-driven structural transformations between the brownmillerite and perovskite phases, with stable intermediate phases. These transitions are accompanied by significant changes in electronic and optical properties [33]. Besides, the analog optical response demonstrated in this work is rooted in the continuous evolution of oxygen stoichiometry and phase coexistence in $SrFeO_{3-\delta}$, as quantitatively established in recent work [34]. By exploiting oxygen-driven structural transformations in a mixed ionic–electronic conductor and combining them with an optically passive interference layer, we achieve continuous and reversible modulation of color and transparency. In this context, we present *ex-situ* and *in-situ* optical measurements, we establish a direct relationship between ion-induced structural evolution, optical constants, and perceptual color control. In addition, we demonstrate that thickness engineering of an optically passive layer enables systematic extension of the accessible color range. These results establish a general framework for ion-driven analog photonic devices and point towards future electrochromic systems.

## 2. Structural, optical and electrochemical characterization of the $SrFeO_{3-\delta}$ \$Al_2O_3$ device.

To exploit these electrically controlled structural and optical transitions, the $SrFeO_{3-\delta}$ / $Al_2O_3$ device illustrated in Figure 1(a) integrates a MIEC thin film with an optically passive dielectric layer to enable structure-driven optical modulation. The schematic presents a vertical multilayer heterostructure in which modulation of oxygen stoichiometry in the MIEC is driven by an applied electrochemical potential across a solid state electrolyte, consisting of a YSZ (100) substrate. A ~10 nm thick Gd-doped ceria (CGO) interlayer was introduced to prevent interfacial reactions and promote epitaxial growth of the 300 nm- $SrFeO_{3-\delta}$ layer.

Ion pumping acts as a continuous control parameter, driving reversible oxygen insertion and extraction within the $SrFeO_{3-\delta}$ layer [35]. Thus, $SrFeO_{3-\delta}$ serves as the optically active medium, while the $Al_2O_3$ capping layer is designed to remain chemically and optically inert. The $Al_2O_3$ capping layer introduces controlled optical interference and prevents oxygen exchange at the top interface. Although $Al_2O_3$ contributes to the overall optical interference conditions, it does not participate in the ion-driven process and does not undergo changes in its optical properties [36,37] . The vertical geometry enables electrical control of oxygen vacancy concentration in the active layer using relatively low voltages, enabling continuous and reversible modulation of the optical response [38].

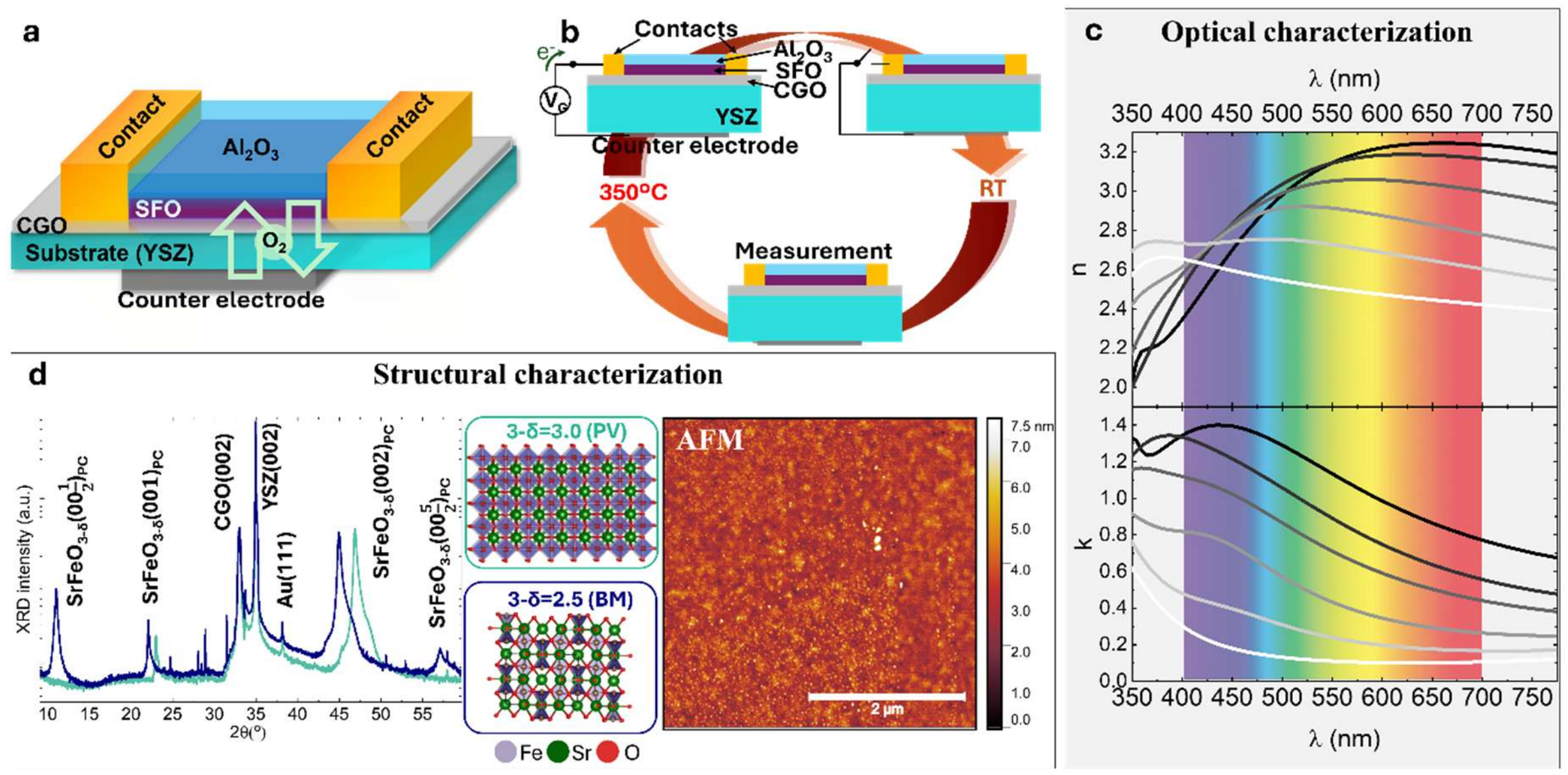

**Figure 1. Device architecture and operating principle.** *(a)* Schematic of the ion-pumped electrochromic device based on a vertical oxide heterostructure. Application of a voltage V drives oxygen ion migration through the YSZ substrate toward the LSFO active layer, inducing reversible changes in its optical properties. *(b)* Cross-sectional image of the device showing the counter electrode, YSZ substrate, CGO interlayer, SFO electrochromic layer, $Al_2O_3$ capping layer, and Au top electrode, confirming the layered architecture and sharp interfaces. *(c)* Optical characterization: (top) refractive index, $n(\lambda)$, and extinction coefficient, $k(\lambda)$, of $SrFeO_{3-\delta}$ with 3-δ from 2 to 2.5. *(d)* Structural characterization: (left) XRD and (center) structure representation of $SrFeO_3$ and $SrFeO_{2.5}$ (right). AFM $Al_2O_3$ after a cycle of reduction and oxidation of the device.

Figure 1(b) schematically shows the operation sequence. In a first step, the device is heated to 350 °C, where YSZ exhibits high oxygen ion conductivity. By applying a voltage, oxygen ions ($O^{2-}$) migrate through the YSZ electrolyte towards or away from the electrochromic, depending on the polarity of the applied bias. Simultaneously, electrons flow through the external circuit to maintain charge neutrality, enabling redox processes in the active layer. After the ion redistribution step, the device is cooled down to room temperature (RT) while maintaining electrochemically the oxygen stoichiometry. At RT, oxygen mobility in YSZ is strongly suppressed, effectively freezing the ionic configuration [39]. This results in a nonvolatile electrochromic state, which can be optically characterized without further ionic motion.

In the final step, optical measurements are performed at room temperature, ensuring that the recorded spectral and colorimetric responses correspond to a stable ionic and structural state of the electrochromic layer. This thermal–electrical cycle enables reversible and reproducible control of oxygen content, phase (brownmillerite ↔ perovskite), and its optical properties using an applied voltage.

The structural evolution of $SrFeO_{3-\delta}$ was investigated using X-ray diffraction (Figure 1b). The fully oxidized perovskite films exhibit diffraction features distinct from those of the oxygen-deficient brownmillerite phase. X-ray diffraction analysis indicates a clear structural evolution of the $SrFeO_{3-\delta}$ phase. The characteristic $SrFeO_{3-\delta}$ $(001/2)_{PC}$ brownmillerite reflection at approximately 11° exhibits significant attenuation, suggesting a reduction in the brownmillerite (BM) phase. Simultaneously, the $(001)_{PC}$ and $(002)_{PC}$ peaks shift from around 22° to approximately 23° and from 44° to about 46°, respectively. These changes are consistent with a transition from the brownmillerite to the perovskite (PV) structure, reflecting a contraction of the lattice parameter associated with oxygen incorporation. Upon ion insertion, the diffraction patterns progressively evolve toward those characteristics of the perovskite structure, indicating increased oxygen content and higher lattice symmetry. Reversing the ion pumping restores the brownmillerite phase, demonstrating the reversibility of the structural transformation and evidencing oxygen stoichiometry modulation within the $SrFeO_{3-\delta}$ lattice. These results confirm that ion insertion induces a controlled and reversible phase transformation in $SrFeO_{3-\delta}$.. The oxygen stoichiometry was quantified by coulometric titration, following the procedure reported for $SrFeO_{3-\delta}$ thin films [31]. During each electrochemical voltage step, the transient current associated with oxygen insertion/extraction was integrated after equilibration, and the resulting charge was converted into a change in oxygen non-stoichiometry using the film volume and unit-cell volume. The most reduced plateau was assigned to the $SrFeO_{2.5}$ brownmillerite state, and the oxygen content of the subsequent states was calculated from the accumulated coulometric charge (see Supporting Information S2). The sharp and well-defined intensity reflections indicate high crystallinity, while atomic force microscopy AFM measurements (Figure 1 (c)) reveal a smooth surface morphology with nanometer-scale roughness after an oxidation and reduction cycle. These results suggest that the optical response is not governed by microstructural evolution of the capping layer but is intrinsically linked to changes in crystalline structure of $SrFeO_{3-\delta}$ .

The ion-driven structural transformation of $SrFeO_{3-\delta}$ leads to pronounced changes in the complex refractive index of $SrFeO_{3-\delta}$. Figure 1(c) presents the refractive index $n(\lambda)$ and extinction coefficient $k(\lambda)$ extracted for representative brownmillerite and perovskite states. These optical constants were obtained by ellipsometry measurements and fitted using a model based on Lorentz oscillations (see Supporting Information Figure S1). The real part (n) increases across the visible range with oxidation, while the extinction coefficient (k) is strongly modulated, particularly in the blue–green range of wavelengths (400-525 nm).

These variations originate from changes in the Fe valence state and the associated electronic band structure, highlighting the strong coupling between ionic configuration and optical response. The continuous evolution of *n* and *k* across the visible spectrum establishes a direct correlation between crystal structure and optical response, providing the material basis for analog optical control. Moreover, the combination of the electrochromic $SrFeO_{3-\delta}$ layer with the optically passive $Al_2O_3$ capping layer introduces interference effects that enhance and tune the perceived color. Thus, $Al_2O_3$ acts as an optically passive layer and that the observed optical modulation originates exclusively from the $SrFeO_{3-\delta}$. layer. The $Al_2O_3$ capping layer plays a critical role by converting these intrinsic material changes into enhanced and tunable optical contrast through interference. Thickness engineering of this passive layer enables systematic control over the spectral position and amplitude of the optical modulation, effectively expanding the range of colors.

## 3. Analog color control

The color range of the device was evaluated through *ex-situ* and *in-situ* reflectance measurements. While *ex-situ* characterization provides high-resolution structural and optical analysis at defined states, *in-situ* measurements enable real-time monitoring of ion transport and optical changes under applied bias, capturing the dynamic processes governing electrochromic switching. In both *ex-situ* and *in-situ* processes different optical states were obtained by electrically tuning the device prior to spectroscopic measurement, applying voltages ranging from -400 to 0 mV at a temperature of 350 ºC to modify the oxygen content and, consequently, the structural phase of the $SrFeO_{3-\delta}$ layer. After each voltage step, the system was measured at RT, enabling direct comparison of nonvolatile states. The spectra were acquired using a setup consisting of a focused white-light source coupled into an optical microscope, with the reflected signal collected and analyzed using a Shamrock SR-30i spectrometer attached to a CCD camera. The incident light was focused onto the active device area, ensuring spatially resolved measurements and minimizing contributions from surrounding regions.

### 3.1 *Ex-situ* color change range

The *ex-situ* acquired reflectance spectra exhibit variations across the visible range (350–750 nm), corresponding to different ionic states in $SrFeO_{3-\delta}$ and a fixed $Al_2O_3$ layer on YSZ substrate induced by applied voltages. We can observe in Figure 2 (a) left, that the spectra show a progressive shift of the dominant reflectance peak from shorter wavelengths (~450 nm) toward longer wavelengths (~600–650 nm) for applied voltages from 0 to -250 mV corresponding to changes in the $SrFeO_{3-\delta}$ structure from perovskite to brownmillerite, accompanied changes in spectral width and intensity. Additionally, a low-intensity, broadband spectrum is observed, corresponding to a quasi-transparent optical state for an applied voltage of -400 mV. These spectral changes directly reflect modifications in both the extinction coefficient (*k*) and the refractive index (*n*) of the $SrFeO_{3-\delta}$ layer, driven by oxygen stoichiometry variations. The interplay between absorption and thin-film interference—enhanced by the $Al_2O_3$ capping layer—results in strong color contrast and tunability.

To translate these spectral responses into perceptual color, the reflectance spectra were converted into CIE 1931 chromaticity coordinates. The resulting points are plotted on the CIE 1931 diagram, see Figure 2 (a) right, where each coordinate corresponds to a perceived color independent of intensity [40]. The device spans a well-defined region within the diagram, indicating a broad but continuous color range. The clustering of points suggests that color tuning occurs within a controlled region of chromatic space, rather than spanning the entire visible spectrum. The transition occurs from blue to gray to orange, followed by transparency at -200 mV, 0 mV, -250 mV, and -400 mV, respectively. Intermediate states are not shown to emphasize the boundaries of the accessible color range.

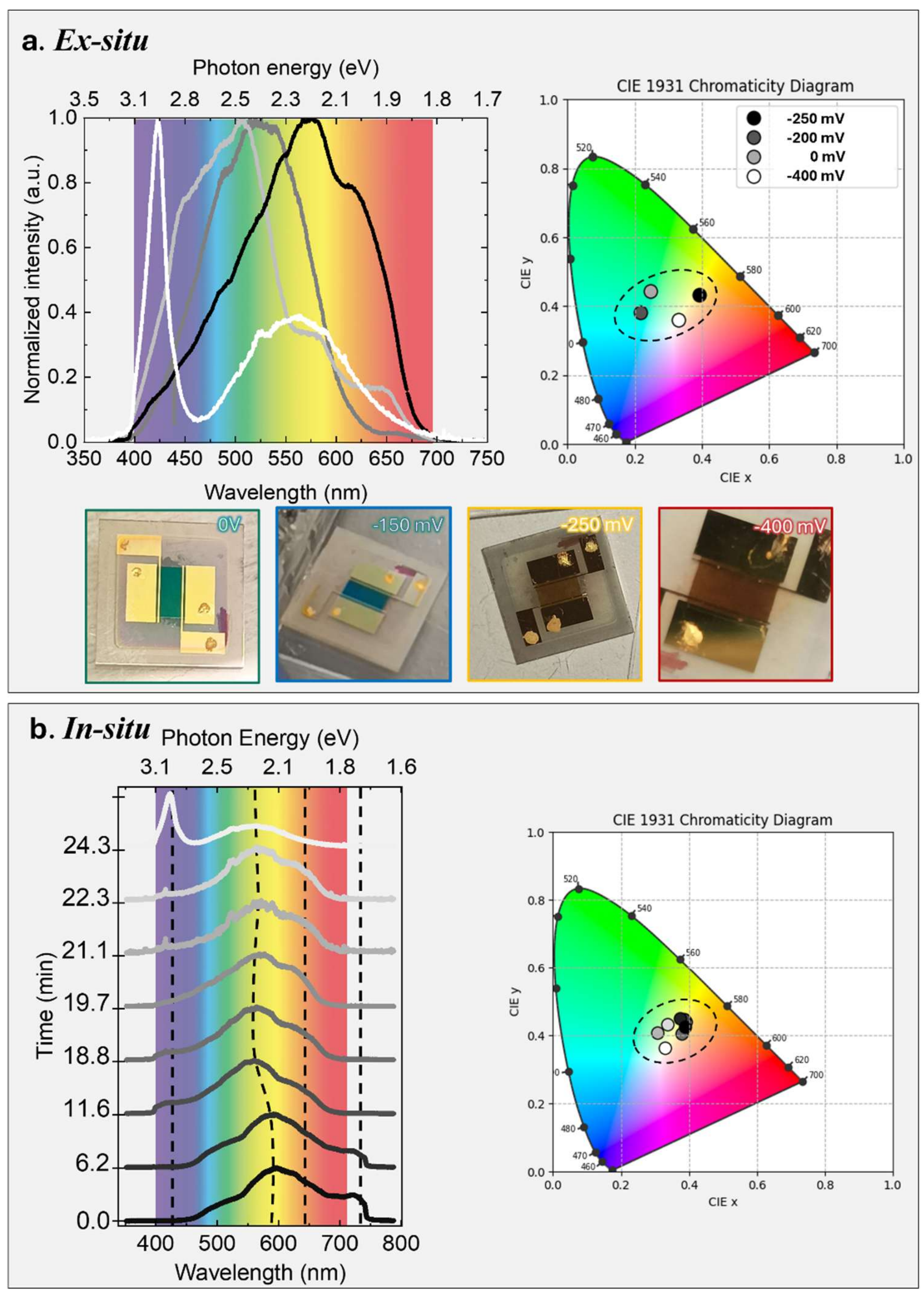

**Figure 2. Analog optical color control in the $SrFeO_{3-\delta}$ / $Al_2O_3$ device.** ***(a)*** *Ex-situ* optical characterization showing normalized reflectance spectra for selected extreme states, spanning a broad spectral range across the visible and its corresponding CIE 1931 chromaticity diagram highlights the accessible color space, including a near-transparent state. Bottom, image of the devices after voltage application of 0, -150, -250 and -400 mV. ***(b)*** *In-situ* optical modulation at a temperature of 350ºC and under applied bias (from 0 to -400 mV). Time-resolved reflectance spectra reveal a continuous spectral evolution and its corresponding CIE chromaticity diagram.

### 3.2 *In-situ* analog color modulation

A dynamic optical response of the device was investigated through *in-situ* measurements under applied bias and an operating temperature of 350ºC (Figure 2(b)). For *in-situ* measurements, the device was placed on a Linkam heating stage, allowing operation at a temperature of 350 °C while applying a voltage in a range between 0 and -400 mV. The incident light was continuously focused onto the active region of the device. The temporal evolution (0–25 min) demonstrates a reproducible response, indicating stable ion dynamics under operating conditions (see Supporting Information, Figure S2). At earlier times, (0–6.2 min), the main reflectance maximum is centered at approximately 590–600 nm, exhibiting only a slight shift toward shorter wavelengths. Between 6.2 and 11.6 min, the spectral feature becomes fastly displaced, shifting to ~575–580 nm. From 11.6 to 18.8 min, this trend continues progressively toward the green–yellow region, reaching approximately 560–570 nm. At advanced times in the process (18.8–24.3 min), the displacement becomes more pronounced, with the dominant feature shifting further to ~540–550 nm, accompanied by a redistribution of spectral intensity toward shorter wavelengths, particularly within the 420–500 nm range. Simultaneously, a redistribution of spectral intensity within the visible range, with an increase observed at 500 nm and a decrease at 650 nm over time. Overall, this behavior is consistent with a continuous blueshift of the reflectance response under applied bias observed in the *ex-situ* measurements.

Notably, the rate of spectral evolution is not uniform across the visible range (see Supporting Information, Figure S3). More pronounced and faster changes are observed in the green–yellow region, where the spectral features shift more rapidly with time, whereas variations at longer wavelengths appear more gradual. This behavior reflects the wavelength-dependent sensitivity of the optical response to changes in the refractive index and extinction coefficient, as well as the interplay with thin-film interference effects. This temporal evolution occurs over characteristic timescales of tens of minutes and is consistent with oxygen-ion transport processes across the oxide heterostructure under applied bias at elevated temperature. In the present device geometry, the response is likely limited by the ionic resistance of the YSZ electrolyte rather than by oxygen diffusion within the $SrFeO_{3-\delta}$ thin film itself, indicating that substantially faster modulation could be reached in fully thin-film solid-state architectures.

The absence of abrupt spectral changes confirms that the process is not governed by discrete phase switching, but rather by a continuous evolution of the oxygen vacancy distribution [41]. The *in-situ* optical modulation is directly reflected in the evolution of the CIE 1931 chromaticity coordinates (Figure 2(b) right). This spectral evolution is directly reflected in the CIE 1931 chromaticity diagram, where the color coordinates trace a continuous trajectory across the dashed region. Specifically, the coordinates evolve from positions associated with more red-shifted (yellow–orange-like) tones toward less red-dominated and slightly greener hues. The smooth displacement of the chromaticity points confirm that the optical modulation proceeds in a continuous and analog manner over time, rather than through abrupt color switching.

Compared to the *ex-situ* measurements, differences are observed both in the reflectance spectra and in the corresponding CIE coordinates. These variations arise from differences in the optical measurement conditions. The presence of the heating stage, specifically its optical window, introduces additional spectral distortions that reduce the measured intensity. Furthermore, slight variations in the focal position during measurement can also influence the collected signal (see Supporting Information, Figure S4). These setup-induced effects lead to minor shifts in the extracted chromaticity coordinates. Although, the continuous evolution of the color is preserved.

## 4. Color range extension by $Al_2O_3$ thickness engineering

### 4.1 Stepped-thickness $Al_2O_3$ design

While the intrinsic electrochromic response of $SrFeO_{3-\delta}$ enables continuous tuning of the optical constants (*n* and *k*), the final color remains limited by the wavelength-dependent spectral response. To overcome this limitation, we introduce an additional degree of freedom based on optical interference engineering via the $Al_2O_3$ capping layer.

To investigate the role of the dielectric capping layer in the optical response, a dedicated device was deposited by PLD consisting of a YSZ substrate with 10 nm-CGO buffer layer and 100 nm $SrFeO_{3-\delta}$ thin film stack but incorporating a lateral $Al_2O_3$ thickness gradient. In this sample, the $Al_2O_3$ capping layer thickness was progressively varied from 375 nm to 325 nm across the device surface. Au current-collector electrodes were patterned on top of the structure to independently address the upper and lower regions of the sample, enabling electrochemical switching between $SrFeO_{2.5}$ (BM) and $SrFeO_3$ (PV) states in spatially separated areas while maintaining the same dielectric thickness gradient. This configuration defines spatially distinct regions with different optical path lengths, while preserving identical electrochemical conditions in the underlying $SrFeO_{3-\delta}$ layer. Consequently, all regions share the same ionic state under bias but exhibit different optical responses due to thickness-dependent phase accumulation upon reflection.

From an optical perspective, the $Al_2O_3$ layer acts as a passive interference medium, introducing wavelength-dependent constructive and destructive interference. The phase shift scales with thickness as $\phi \propto n_{Al_2O_3} \cdot d/\lambda$, ($n_{Al_2O_3}$ as the refractive index of $Al_2O_3$ , d the capping layer thickness and $\lambda$ the wavelength) enabling precise control of the spectral position of reflectance maxima and minima [42]. This approach effectively decouples the active layer (ionic modulation) from the passive layer which allows tunning the range of colors, allowing independent optimization of both.

The impact of thickness variation on the optical response can be observed in the simulated reflectance spectra performed by IMD program (Figure 3(b) center) [43]. Each thickness region exhibits a different spectral profile, with interference shifting across the visible range. Specifically, 325 nm thick $Al_2O_3$ layers produce reflectance maxima at shorter wavelengths (blue–green region), while increasing thickness of 350 and 375 nm induces a progressive redshift toward longer wavelengths (blue and yellow–red region, respectively). This behavior arises from the increase in optical path length within the capping layer, which modifies the interference condition for reflected light. The spectral modulation is not limited to a rigid shift but also involves changes in spectral contrast and bandwidth when $SrFeO_{3-\delta}$ changes its structure modulating n and k, reflecting the complex interplay between interference and the wavelength-dependent optical constants of $SrFeO_{3-\delta}$.

The thickness simulation corresponding to the CIE chromaticity coordinates (Figure 3(b) right) reveal that each thickness region occupies a distinct position in color coordinates. Compared to the uniform-thickness device, the variable thickness architecture significantly expands the accessible chromatic range, enabling coverage of a broader region within the CIE diagram. Notably, this color diversification is achieved without altering the material composition or the ionic state, but purely through geometric control of optical interference.

### 4.1 Simulation of optical properties of $SrFeO_{3-\delta}$

Figure 3(c) establishes the direct link between the intrinsic optical properties of $SrFeO_{3-\delta}$, with its structure, and the resulting perceptual color. Figure 3(c) left, shows the experimentally measured optical absorption coefficient *(α)* of $SrFeO_{3-\delta}$ across different oxygen stoichiometries, corresponding to the structural evolution from the brownmillerite (BM) phase to the perovskite (PV) phase, including intermediate states. The optical absorption coefficient (α) exhibits a clear voltage-dependent evolution across the visible range. As the applied voltage is varied from positive to negative values, the overall absorption decreases progressively, with the highest absorption observed at positive bias (75 mV) and the lowest at more negative voltages ( −400 mV). This reduction in α is accompanied by a gradual flattening of the spectral profile, particularly in the 400–600 nm range, indicating a modification of the electronic structure of $SrFeO_{3-\delta}$. In addition to the overall decrease in magnitude, subtle changes in the spectral slope are observed, reflecting a continuous evolution of the electronic transitions associated with oxygen stoichiometry. This behavior confirms that the applied voltage effectively tunes the absorption properties of the material, providing the basis for the observed modulation of the reflectance spectra.

Based on these optical constants, ellipsometry-derived simulations were performed to model the optical response of the full $Al_2O_3$/$SrFeO_{3-\delta}$/CGO/YSZ stack. These simulations capture the transition of reflectance spectra at an angle of 0°, as the material evolves from the BM to the PV phase, reproducing the gradual spectral shifts and changes in intensity observed experimentally. The effect of the angle of incidence on the chromaticity is shown in the Supporting Information (Figure S5).

The simulated reflectance spectra in Figure 3(c) center exhibits a clear voltage-dependent shift of the main interference peak across the visible range. As the applied voltage is varied from positive to negative values, the dominant reflectance maximum progressively shifts from shorter to longer wavelengths, moving from approximately ~480–500 nm at higher voltages (75 mV) toward ~550–600 nm at more negative voltages (−400 mV). This redshift is accompanied by changes in peak intensity and spectral shape, indicating a strong dependence of the optical response on the underlying oxygen stoichiometry. This behavior reflects the continuous evolution of the optical constants (n and k) of $SrFeO_{3-\delta}$ as a function of the applied voltage, which modifies both the absorption profile and the interference conditions within the multilayer stack. The smooth and progressive displacement of the spectral feature confirms that voltage acts as an effective control parameter for tuning the optical response.

To translate these spectral changes into perceptual color, the simulated reflectance spectra of the multilayer stack was projected onto the CIE 1931 chromaticity diagram (Figure 3(c) right). The resulting coordinates follow a continuous trajectory, demonstrating how structural evolution at the atomic scale maps directly onto color space. This analysis highlights that, while the optical constants ($n$, $k$) of $SrFeO_{3-\delta}$ can be continuously tuned through ionic control, the resulting color space remains limited by their wavelength-dependent dispersion. While ionic control influences the spectral response, it alone does not facilitate a broad chromatic range without additional optical engineering improvements.

Small variations in the angle of incidence lead to minor changes in the spectral response and, consequently, in the perceived color, although the overall chromatic range remains preserved. The results in Figure 3(c) confirm that the structural transformation directly governs the spectral response through modifications of the optical properties.

Figure 3d shows the impact of the $Al_2O_3$ capping layer thickness on the optical response of the device, highlighting its role as a passive interference element for color tuning. The optical image of the device (Figure 3d, left) provides direct visual confirmation of this effect, showing spatially resolved color variation across the stepped-thickness regions and between electrochemically switched areas. In Figure 3 (d) left, an image of the device reveals two main regions, top and bottom with $SrFeO_{2.5}$ (BM) and $SrFeO_3$ (PV), respectively, in both parts $Al_2O_3$ thickness decrease from left (375 nm) to right (325 nm). A clear gradient of colors is observed across the sample in the bottom part, and from bottom to top ranging from strongly colored regions to a nearly transparent state. Notably, the region labeled “4” appears almost transparent, consistent with a brownmillerite-like state with low reflectance contrast. In contrast, the colored regions, separated by the Au contacts, are numbered 1,2,3 corresponding to the perovskite phase thin film with thicknesses of 375, 350 and 325 nm, respectively. The transition between the colored and transparent regions is spatially well-defined, with a sharp boundary separating the transparent (brownmillerite, off state) and colored (perovskite, on state) areas.

In Figure 3 (d) center, spectrometer spectra acquired from each numbered region are presented. These measurements were performed locally, probing each thickness-defined area independently and the underlying $SrFeO_{2.5}$ (BM) and $SrFeO_3$ (PV) structure. For the $SrFeO_3$ (PV) structure, spectra exhibit a shift of the main peak towards higher wavelengths (from 500 to 575 nm) as a function of decreasing $Al_2O_3$ thickness. At the same time, the overall spectral shape is influenced by the optical constants of $SrFeO_{3-\delta}$, then, $SrFeO_{2.5}$ (BM) spectra shows a double intensity peak with two maxima at 450 and 600nm. In Figure 3(d) right, each experimentally measured spectrum is converted into CIE 1931 chromaticity coordinates, enabling direct visualization of the corresponding perceptual color. The distribution of points confirms that thickness engineering of the capping layer enables access to a broader color space compared to the response of the material alone.

Together, these results demonstrate that the $Al_2O_3$ layer acts as a tunable optical cavity, enabling precise control of the spectral distribution of reflected light. When combined with the continuous tuning of optical constants in $SrFeO_{3-\delta}$, this approach allows decoupled and synergistic control of color, establishing thickness engineering as a powerful strategy to extend the functionality of ionochromic photonic devices.

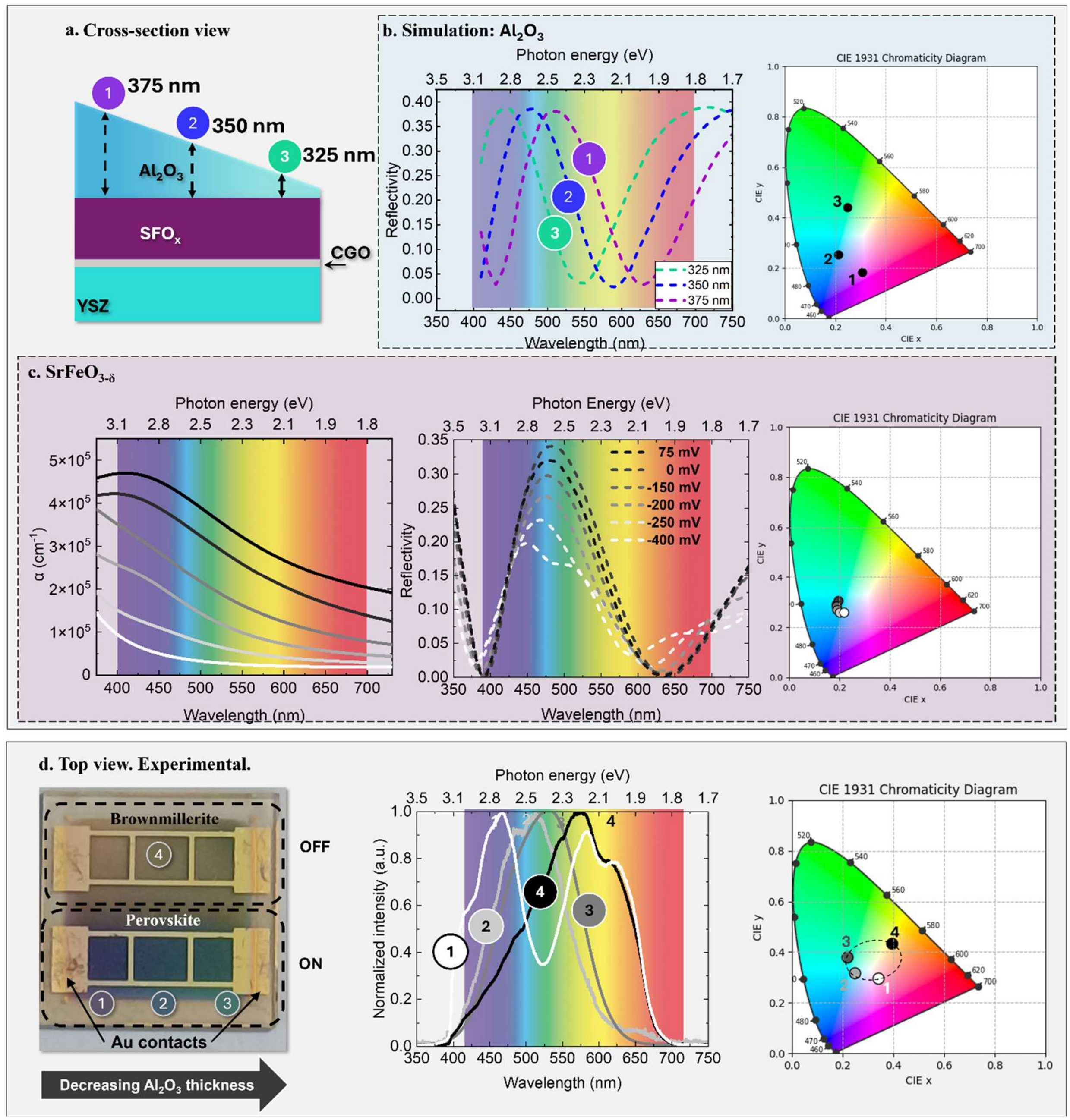

**Figure 3. Color range extension via $Al_2O_3$ thickness engineering. (a)** Cross-sectional schematic of the device showing a stepped $Al_2O_3$ capping layer with increasing thickness, defining regions with different optical path lengths. **(b)** Optical simulations of reflectance for different $Al_2O_3$ thicknesses, demonstrating thickness-dependent interference modulation across the visible spectrum, together with the corresponding CIE chromaticity coordinates. **(c)** Evolution of the optical absorption coefficient and reflectance of $SrFeO_{3-\delta}$ under different applied voltages. **(d)** Experimental demonstration of thickness-engineered color tuning. Left: optical image of the device showing distinct color regions associated with different

$Al_2O_3$ thicknesses. Center: corresponding reflectance spectra. Right: extracted CIE chromaticity coordinates, confirming the extension of the accessible color gamut through interference engineering.

**Conclusions**

In this work, we demonstrate structure-driven analog optical control in ion-pumped $SrFeO_{3-\delta}$ thin-film devices, establishing a direct relationship between oxygen stoichiometry, structural phase, optical constants, and perceptual color. By electrically modulating the oxygen content using applied voltages in the range of 0 to −400 mV, reversible transformations between BM and PV phases are achieved, enabling continuous and nonvolatile tuning of the optical response.

*Ex-situ* reflectance measurements reveal a broad and well-defined color range associated with distinct ionic states, with the dominant reflectance peak shifting from ~450 nm to ~600–650 nm across the accessible voltage window. *In-situ* experiments at 350 °C provide direct evidence of a gradual and time-dependent spectral evolution, occurring over a characteristic timescale of ~25 min. During this process, the main reflectance feature progressively shifts from ~590–600 nm to ~540–550 nm, consistent with a continuous blueshift driven by oxygen ion migration. The absence of abrupt spectral transitions, together with the continuous displacement of the CIE chromaticity coordinates, confirms that the device operates in a true analog regime, where intermediate optical states can be reliably accessed and stabilized.

Furthermore, we show that the optical response is not solely determined by the intrinsic properties of $SrFeO_{3-\delta}$ but can be actively engineered through thin-film interference. By introducing a thickness-controlled $Al_2O_3$ capping layer in the range of ~325–375 nm, we decouple ionic modulation from optical interference effects, enabling systematic tuning of the spectral response. This approach allows the expansion of the accessible color space beyond the limits imposed by the material dispersion alone.

Overall, this work demonstrates that oxygen-driven ionic control can act as a continuous tuning parameter linking structural transformations to optical functionality. The combination of ion-controlled optical constants with interference engineering provides a versatile platform for designing reconfigurable photonic devices. These results open new opportunities for the development of adaptive optical systems, ionochromic displays, and neuromorphic photonic architectures compatible with oxide electronics and silicon-based technologies.

# Methods

## Sample fabrication

Epitaxial $SrFeO_{3-\delta}$ thin films were grown on (001)-oriented YSZ substrates (1 × 1 $cm^2$) with a gadolinium-doped ceria (CGO) buffer layer. Deposition was carried out by pulsed laser deposition (PLD) using a large-area system (PVD Products) equipped with a KrF excimer laser (λ = 248 nm, Lambda Physik). A ~10 nm CGO layer was first deposited to prevent interfacial reactions between $SrFeO_{3-\delta}$ and YSZ. Subsequently, $SrFeO_{3-\delta}$ films with different thicknesses were grown under identical conditions. During all depositions, the substrate temperature was maintained at 600 °C under an oxygen partial pressure of 0.0067 mbar, with a target–substrate distance of 90 mm. The laser fluence was 0.8 J·$cm^{-2}$ and the repetition rate was 10 Hz. Electrical contacts (10 nm Ti / 100 nm Au) were patterned by photolithography followed by thermal evaporation. Finally, an $Al_2O_3$ capping layer (~350 nm) was deposited by PLD under the same conditions. This layer was patterned to leave parts of the Au electrodes exposed for electrical probing. Silver paste was applied on the backside of the YSZ substrate to serve as a counter electrode. After each deposition step, thickness and roughness were measured to ensure precise control of the film parameters.

## Electrochemical characterization

The electrochemical state of the films was controlled by applying different working voltages ($V_W$), which shift the oxygen equilibrium in $SrFeO_{3-\delta}$. This effect can be described as a change in the equilibrium oxygen partial pressure ($p_{O_2}^{eq}$) following a Nernst-type relation:

$$p_{O_2}^{(eq)} = p_{O_2}^{(ref)} \cdot exp\left(\frac{4e\Delta V}{k_B T}\right)$$

where $p_{O_2}^{ref}$ is the ambient oxygen partial pressure.

Changes in oxygen stoichiometry generate a transient current ($I_W$) that decays as equilibrium is reached. The total transferred charge, obtained by integrating the current overtime, is directly related to the variation in oxygen non-stoichiometry (Δδ):

$$\Delta\delta = -\left(\frac{V_{uc}}{2eV_{film}}\right)\int Iw(t)\,dt$$

where $e$ is the elementary charge, $V_{uc}$ is the unit cell volume, and $V_{film}$ is the film volume.

**X-Ray diffraction measurements**

Structural evolution during electrochemical operation was studied by *in-situ* X-ray diffraction using a heating stage (Anton Paar DCS 1100) combined with a Keithley 2401 source meter. Electrical connections were established using three platinum wires: one connected to the counter electrode via the silver-coated stage holder, and two connected to the top electrodes using gold paste. The sample alignment was optimized by adjusting the stage height and tilt relative to the X-ray beam. Measurements were performed at 325 °C in ambient air. Two voltages −0.45 V and 0.1 V were applied. XRD patterns were acquired after current stabilization.

**Ellipsometry measurements**

Optical properties were characterized by spectroscopic ellipsometry using a Horiba UVISEL system. Measurements were performed in the energy range from 0.6 to 5.0 eV (step size: 0.05 eV) at an incidence angle of 70°. Data analysis was carried out using DeltaPsi2 software. During measurements, voltages between −0.40 V and 0.075 V were applied. *In-situ* spectra were acquired after stabilization at each voltage step.

**CIE 1931 Chromaticity Diagram**

The chromaticity coordinates were obtained from the measured spectra using a custom Python script developed for this work. The routine imports the spectral intensity data, interpolates each spectrum to the 380–780 nm range with 1 nm resolution, and converts the resulting spectral distributions into CIE 1931 tristimulus values, chromaticity coordinates, and sRGB values using the *colour-science* library. The chromaticity points were then plotted on the CIE 1931 diagram using *Matplotlib*.

The conversion from the spectral intensity distribution $I(\lambda)$to the CIE XYZ tristimulus values is given by:

$$X = \int_{380}^{780} I(\lambda)\ \bar{x}(\lambda)\, d\lambda$$
$$Y = \int_{380}^{780} I(\lambda)\ \bar{y}(\lambda)\, d\lambda$$
$$Z = \int_{380}^{780} I(\lambda)\ \bar{z}(\lambda)\, d\lambda$$

where $\bar{x}(\lambda)$, $\bar{y}(\lambda)$, and $\bar{z}(\lambda)$are the CIE 1931 color matching functions.

The chromaticity coordinates $(x, y)$are then obtained as:

$$x = \frac{X}{X + Y + Z}, y = \frac{Y}{X + Y + Z}$$

Finally, the XYZ values are converted into sRGB color space through a linear transformation followed by gamma correction:

$$\begin{bmatrix} R \\ G \\ B \end{bmatrix} = \mathbf{M} \begin{bmatrix} X \\ Y \\ Z \end{bmatrix}$$

where $\mathbf{M}$is the standard XYZ-to-sRGB transformation matrix. The resulting RGB values are fixed to the [0,1] range for visualization.

# Supporting Information

**Structure-driven analog optical control in ion-pumped $SrFeO_{3-\delta}$ thin-film devices**

Alicia Ruiz-Caridad[1], Paul Nizet[1], Francesco Chiabrera[1], Xavier Vea[1], Philipp Langner[1], Alex Morata[1], Albert Tarancon[1].

*Department of Advanced Materials for Energy Catalonia Institute for Energy Research (IREC) Jardin de les Dones de Negre 1, Sant Adrià de Besòs (Barcelona) 08930, Spain*

### S1. Ellipsometry modelling

The spectroscopic ellipsometry measurements were analyzed by fitting the experimental Ψ and Δ spectra using the DeltaPsi2 software package. The optical response of the SFO thin films was described through a dielectric function constructed as a sum of Lorentz oscillators, expressed as:

$$\varepsilon(\omega) = \varepsilon_\infty + \sum_{j=1}^{n} \frac{f_j \omega_{0,j}^2}{\omega_{0,j}^2 - \omega^2 + i\gamma_j \omega}$$

In this expression, $\varepsilon_\infty$ represents the high-frequency dielectric constant, while $\omega$corresponds to the angular frequency of the incident light. The parameters $\omega_{0,j}$, $f_j$, and $\gamma_j$ denote the resonance frequency, oscillator strength, and damping factor of each oscillator, respectively.

The resulting dielectric function was used to extract the refractive index (n) and extinction coefficient (k) that reproduce the experimental ellipsometry spectra across different oxygen stoichiometries. The fitting procedure considered a multilayer structure consisting of YSZ, CGO, the SFO film, an $Al_2O_3$ capping layer, and a surface roughness layer approximated as an effective medium composed of equal fractions of $Al_2O_3$ and vide.

The thickness of each layer was fixed based on atomic force microscopy measurements. During the fitting process, only the oscillator parameters associated with the SFO layer were treated as adjustable variables, while the optical constants of the remaining layers were independently determined over a range of temperatures.

Uncertainties in the extracted optical conductivity were estimated from the sensitivity of the fitted oscillator parameters provided by the DeltaPsi2 software. The optical conductivity was calculated as:

$$\sigma(\omega) = \frac{4\pi nk}{\lambda Z_0}$$

where $Z_0 = \sqrt{\mu_0/\varepsilon_0} = 377\ \Omega$ is the impedance of free space. This quantity represents the effective electrical response of the material under an alternating electromagnetic field at angular frequency $\omega$.

**S2. Electrochromic response under voltage bias**

To investigate the electrochromic switching dynamics, time-resolved electrical measurements were performed under different applied voltages. The current response was recorded during voltage steps, providing insight into the dynamic of ionic transport and associated redox processes within the $SrFeO_{3-\delta}$ layer. As shown in Figure S3, the current exhibits a characteristic transient response, with an initial sharp peak followed by a gradual decay toward a steady-state. This behavior is consistent with capacitive charging and fast initial ionic redistribution, followed by slower diffusion-limited processes governing oxygen vacancy migration. Higher voltages result in stronger initial currents and faster relaxation, reflecting enhanced ionic mobility. The reproducibility of these transients across multiple voltage steps confirms the stability of the device functioning.

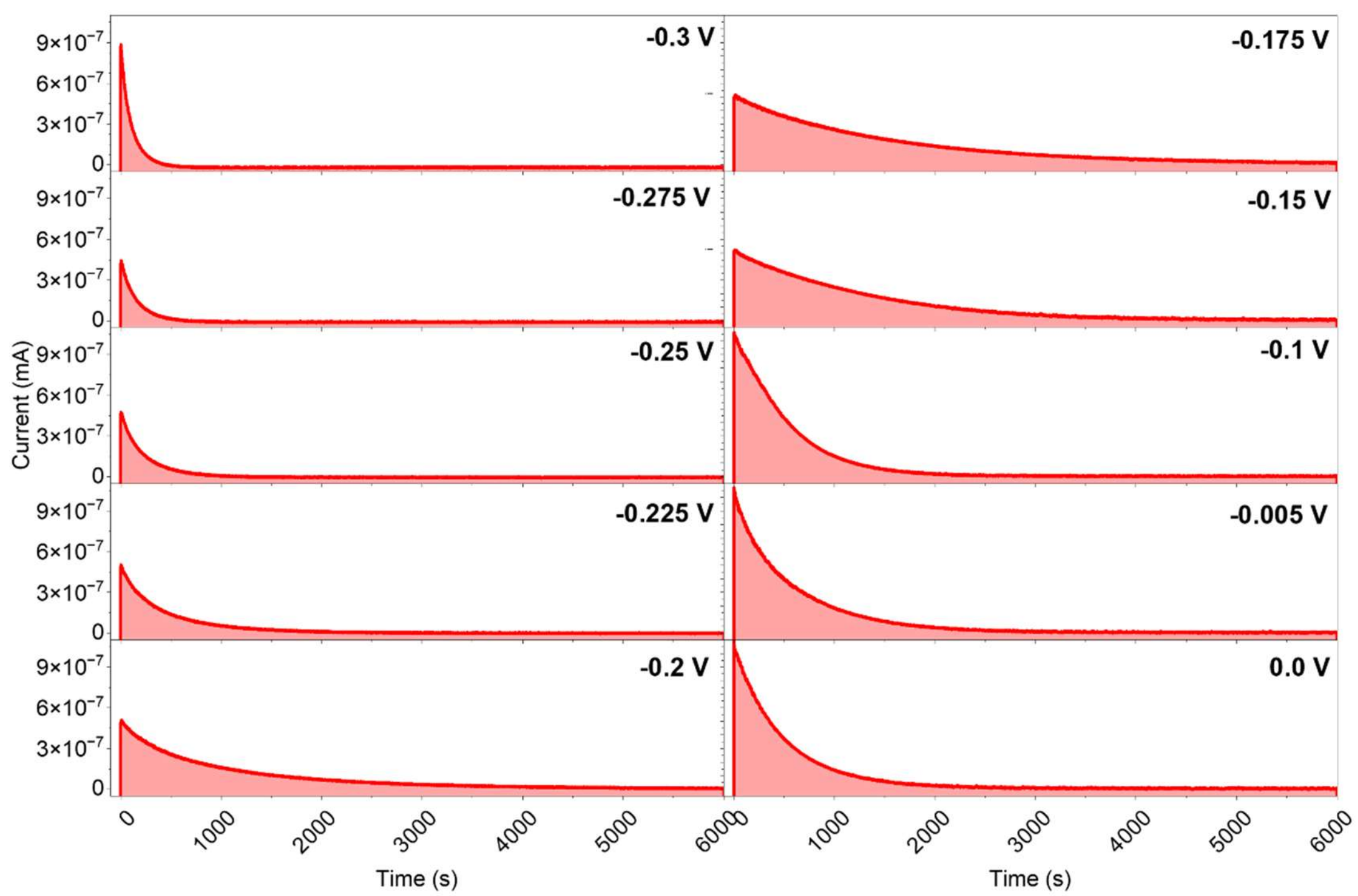

**Figure S2. Time-resolved electrical response of the device under different applied voltages.** Current response of the $SrFeO_{3-\delta}/Al_2O_3$ device under successive voltage steps ranging from -0.3 V to 0 V.

The switching dyanmics between different electrochromic states were further analyzed by monitoring the time-dependent current response during voltage-induced transitions. As shown in Figure S3, the current decay follows a non-instantaneous relaxation, indicating that the switching process is governed by ion migration and diffusion within the $SrFeO_{3-\delta}$ layer. Initially, a rapid decrease in current is observed, corresponding to the immediate response of mobile charge carriers. This is followed by a slower relaxation regime, associated with the redistribution of oxygen vacancies across the film thickness.

**S3. Influence of the focus on the spectroscopy measurements**

To assess the spatial sensitivity of the optical measurements, reflectance spectra were acquired at different focal positions across the device. By adjusting the focal plane along the vertical axis of the multilayer structure, different effective probing volumes were sampled, including regions dominated by the $Al_2O_3$ capping layer, the $SrFeO_{3-\delta}$ active layer, and their optical interference region.

The reflectance spectra exhibit variations depending on the focal position. The relative intensity and position of the spectral maxima shift as the focus is moved, reflecting changes in the effective optical path and the relative contribution of each layer to the measured signal. When the focus is closer to the surface ($Al_2O_3$ region), interference effects are more pronounced. It is observed that a maximum intensity is placed in the blue range of wavelengths. When the focus deepens, the contribution of the blue wavelengths is attenuated and a local maximum in the yellow-red wavelengths arises, enhancing the contribution of the $SrFeO_{3-\delta}$ layer, where absorption and electronic structure dominate the response. These variations are further reflected in the corresponding CIE chromaticity coordinates (Figure S2b), where each focal position maps to a shift point in color space. Although the overall color remains within the same region of the chromaticity diagram, a measurable spread is observed, indicating that the perceived color is sensitive to the exact measurement conditions

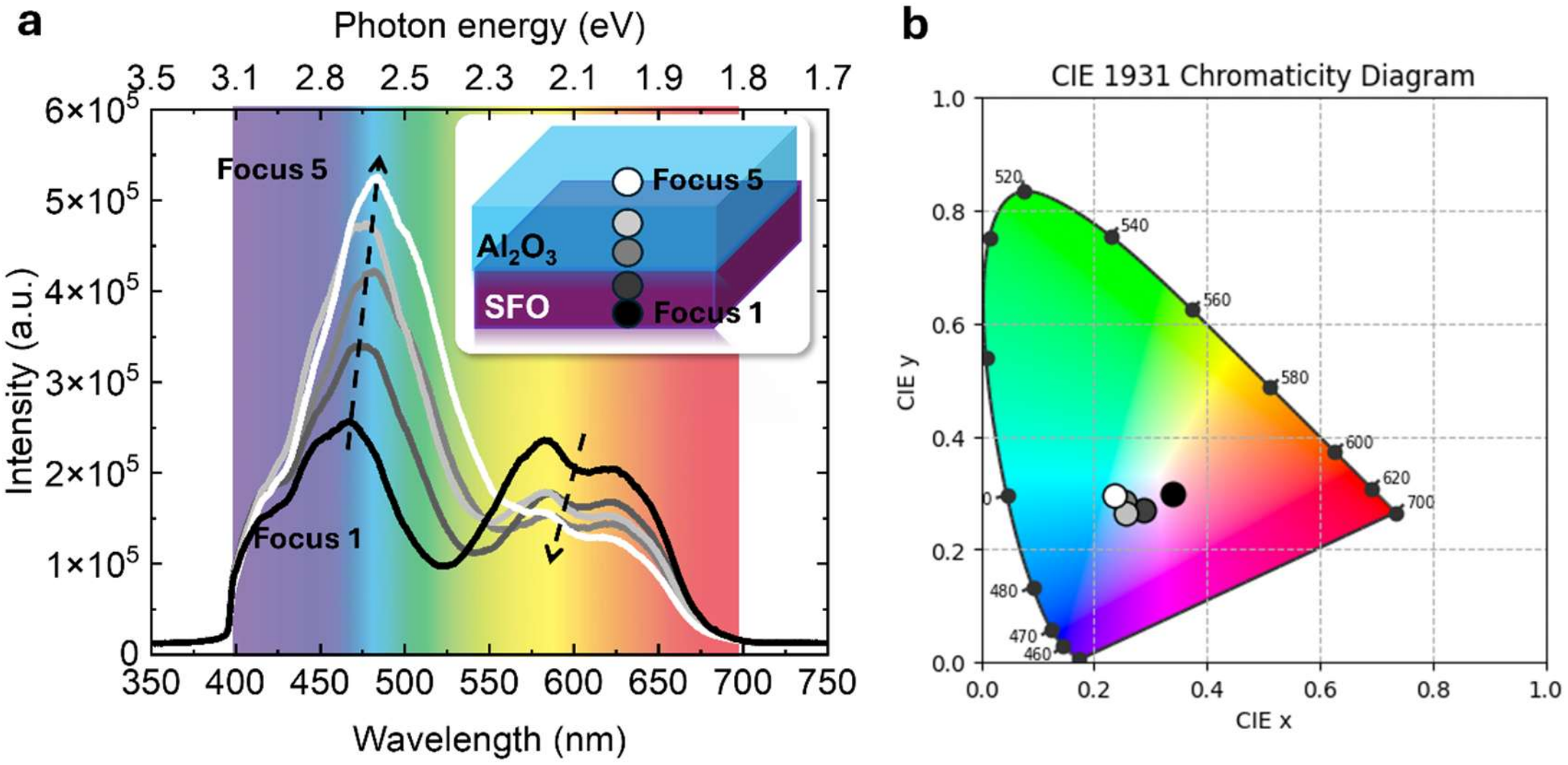

**Figure S3. Influence of measurement focus on reflectance spectra. (a)** Reflectance spectra acquired at different focal positions across the $SrFeO_{3-\delta}$ / $Al_2O_3$ device. The schematic inset indicates the relative focus positions across the $Al_2O_3$ capping layer and the underlying $SrFeO_{3-\delta}$ film. **(b)**The corresponding CIE 1931 chromaticity coordinates show the resulting dispersion in perceived color, highlighting the sensitivity of optical measurements to the exact probing location.

**S4. Time evolution of *in-situ* electrochromic state transitions.**

Switching time exhibits a strong dependence on the applied voltage, reflecting the underlying structural evolution of the $SrFeO_{3-\delta}$ layer from the brownmillerite ($SrFeO_{2.5}$, BM) to the perovskite ($SrFeO_3$, PV) phase. A clear change in behavior is observed near $3–\delta \approx 2.7$, marking the onset of phase coexistence. Three distinct regimes can be identified from the slope of the temporal response. The initial transition from the BM phase to the intermediate phase (IPh) proceeds relatively slowly, followed by a faster switching within the intermediate regime, and a subsequent slowdown as the system approaches the PV phase. This behavior is attributed to the different kinetics of oxygen incorporation associated with each structural regime. These results confirm that the electrochromic response of the device is not instantaneous but occurs over well-defined and controllable timescales. The smooth and reproducible temporal evolution further supports the analog nature of the switching process, where intermediate states can be accessed continuously rather than through discrete transitions.

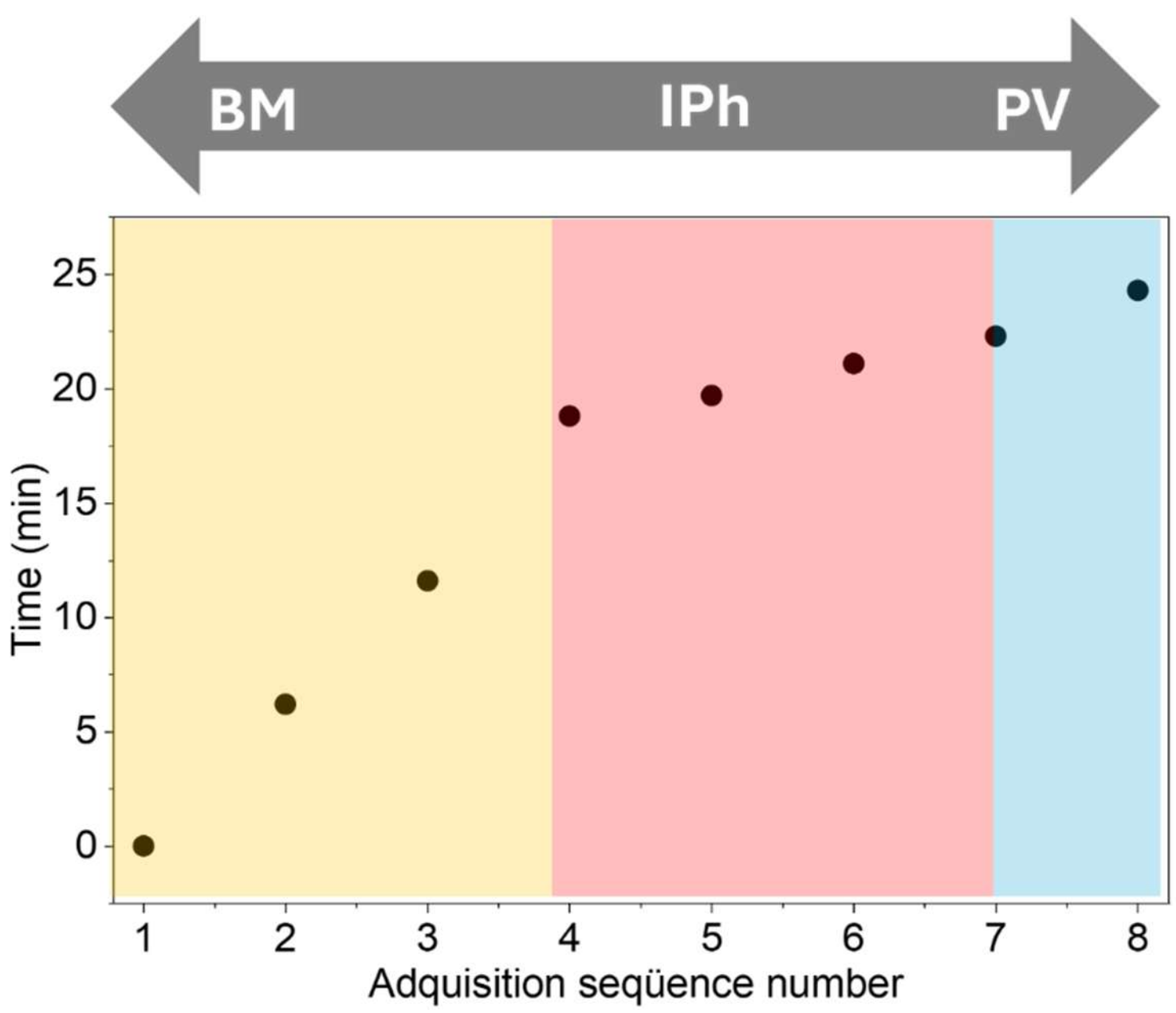

**Figure S4. Dynamics of electrochromic switching between different ionic states.** Time evolution of the transitions between electrochromic states under applied bias.

**S5. Incident angle simulations based on ellipsometry measurements.**

To evaluate the influence of observation geometry on the optical response, angle-dependent reflectance spectra were simulated using the optical constants (*n*, *k*) extracted from ellipsometry measurements of the $SrFeO_{3-\delta}$ layer. As shown in Figure S1 (top left), the reflectance spectra exhibit a clear dependence on the angle of incidence. Increasing the angle from 0° to 30° results in a systematic shift of the spectral features, particularly in the visible range. This behavior originates from the angle-dependent optical path length, which modifies the phase accumulation of reflected light and, consequently, the conditions for constructive and destructive interference. The impact of these spectral changes on perceptual color is captured by projecting the simulated spectra onto the CIE 1931 chromaticity diagram (Figure S1, top right and bottom images). Each angle corresponds to a distinct cluster of chromaticity coordinates, revealing a measurable shift in perceived color with viewing angle. Despite these variations, the color points remain confined within a relatively narrow region of the chromatic space, indicating that the device maintains a consistent color response with moderate angular dependence.

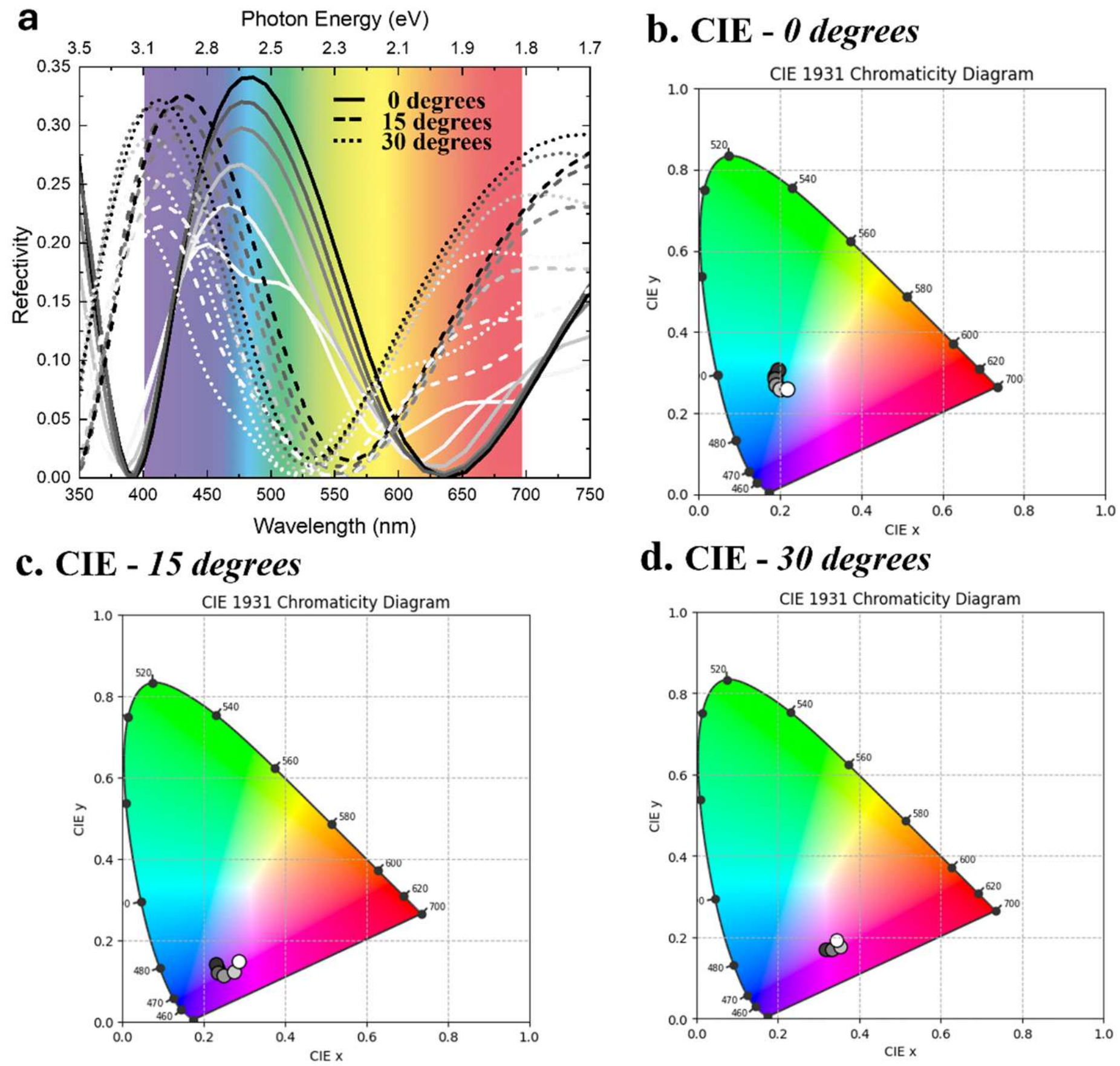

**Figure S5. Angle-dependent optical response and color coordinates derived from ellipsometry simulations.** Simulated reflectance spectra of the $SrFeO_{3-\delta}$ structure at different angles of incidence (0°, 15°, and 30°), based on simulated ellipsometry extracted optical constants (top left). The spectral evolution reveals angle-dependent interference effects, resulting in shifts of the reflectance features across the visible range. The corresponding CIE 1931 chromaticity coordinates are shown for each angle (0°-top right, 15°- bottom right and 30°- left).